\documentclass[twocolumn,showpacs,preprintnumbers,amsmath,amssymb,superscriptaddress]{revtex4}

\usepackage{graphicx}% Include figure files
\usepackage{dcolumn}% Align table columns on decimal point
\usepackage{bm}% bold math

\begin{document}

\title{Influence of thermal environment on optimal working conditions of thermoelectric generators}

\author{Y. Apertet}\email{yann.apertet@gmail.com}
\affiliation{Lyc\'ee Jacques Pr\'evert, F-27500 Pont-Audemer, France}
\affiliation{Institut d'Electronique Fondamentale, Universit\'e Paris-Sud, CNRS, UMR 8622, F-91405 Orsay, France}
\author{H. Ouerdane}
\affiliation{Russian Quantum Center, 100 Novaya Street, Skolkovo, Moscow Region 143025, Russia}
\affiliation{Laboratoire Interdisciplinaire des Energies de Demain (LIED), CNRS UMR 8236, Universit\'e Paris Diderot, 5 Rue Thomas Mann, 75013 Paris, France}
\author{C. Goupil}
\affiliation{Laboratoire Interdisciplinaire des Energies de Demain (LIED), CNRS UMR 8236, Universit\'e Paris Diderot, 5 Rue Thomas Mann, 75013 Paris, France}
\author{Ph. Lecoeur}
\affiliation{Institut d'Electronique Fondamentale, Universit\'e Paris-Sud, CNRS, UMR 8622, F-91405 Orsay, France}

\date{\today}

\begin{abstract}
Optimization analyses of thermoelectric generators operation is of importance both for practical applications and theoretical considerations. Depending on the desired goal, two different strategies are possible to achieve high performance: through optimization one may seek either power output maximization or conversion efficiency maximization. Recent literature reveals the persistent flawed notion that these two optimal working conditions may be achieved simultaneously. In this article, we lift all source of confusion by correctly posing the problem and solving it. We assume and discuss two possibilities for  the environment of the generator to govern its operation: constant incoming heat flux, and constant temperature difference between the heat reservoirs. We demonstrate that, while power and efficiency are maximized simultaneously if the first assumption is considered, this is not possible with the second assumption. This latter corresponds to the seminal analyses of Ioffe who put forth and stressed the importance of the thermoelectric figure of merit $ZT$. We also provide a simple procedure to determine the different optimal design parameters of a thermoelectric generator connected to heat reservoirs through thermal contacts with a finite and fixed thermal conductance.
\end{abstract}


\maketitle
\section{Introduction}

Thermoelectric generators provide a convenient way to convert a heat flux into electrical power since their operation does not involve moving parts, and they may be used to design small systems without loss of efficiency due to size effects, contrary to classical heat engines \cite{Vining2009}. However, the performances of thermoelectric generators still remain insufficient to envisage yet a massive development and use of this technology. To circumvent this obstacle, research mainly focuses on the improvement of the thermoelectric properties of materials (see, e.g., Ref.~\cite{Snyder2008} and references therein). Owing to intensive investigations and use of nanostructuration, the best thermoelectric figure of merit $Z\overline{T}$ has doubled over the last decade, and now reaches values around 2 \cite{Heremans2013}. In parallel to materials research, there is also a need for a reflection on the working conditions of TEGs, and how to optimize these. Ioffe in his seminal work~\cite{Ioffe1957} clearly gives working conditions for maximum efficiency, and maximum output power in the simple case of a TEG with constant thermoelectric parameters. Since then, these conditions have also been derived for various refined models \cite{Handbook}. While efficiency and output power are maximized for different working conditions in Ioffe's model system, the possibility to reach \emph{simultaneously} both optimal working conditions was recently discussed \cite{Baranowski2013, Baranowski2014}. In this article we demonstrate that this requires very specific assumptions on the thermal flux, quite different from those made by Ioffe~\cite{Ioffe1957}. Our purpose here is to clearly distinguish these two different sets of assumptions in order to avoid confusion in debates concerning power maximization, which is the most relevant optimization target from the practical viewpoint.

The article is organized as follows. In Sec.~\ref{comparison}, we discuss the two possible assumptions on the thermal environment of the TEG. Since the case of constant temperature difference between thermal reservoirs is well known, we then focus on the case where the incoming thermal flux is constant. In particular, we derive the working condition leading to both efficiency and power maximization. We also discuss recent literature in light of our results. Then, in Sec.~\ref{procedure}, we turn to practical concerns on power maximization of a TEG and we propose a simple optimization procedure to determine the optimal design parameters of a TEG placed in a given thermal environment.

\section{\label{comparison} A tale of two assumptions}

Optimal working conditions of a TEG depend on its thermal environment; so the choice of the thermal constraints imposed on the TEG naturally plays an important role in the derivation of these conditions. We demonstrate here that choosing a constant temperature difference between the heat reservoirs or a constant incoming flux leads to quite different optimization strategies.

\subsection{Description of the thermoelectric system}

\begin{figure}
	\centering
		\includegraphics*[width=0.47\textwidth]{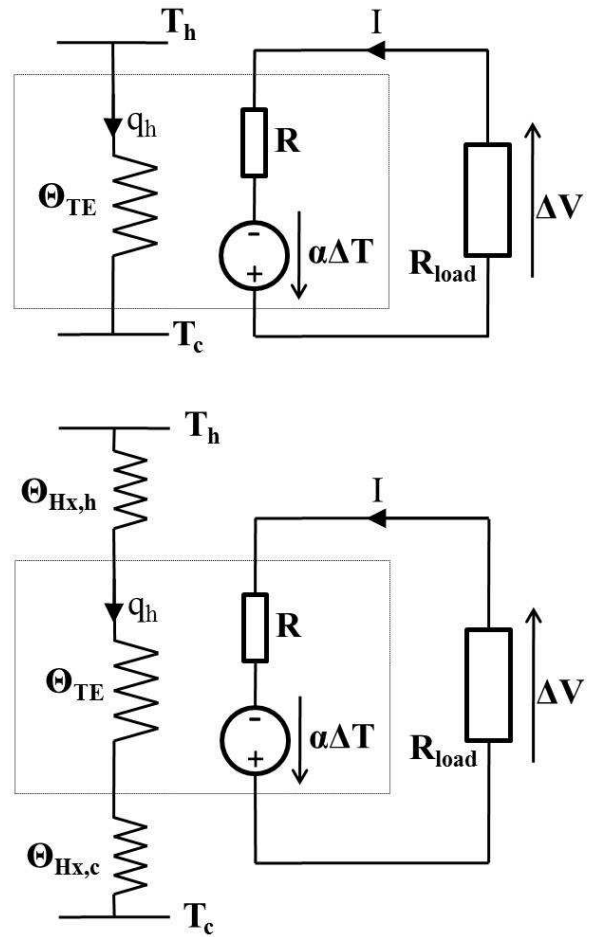}
	\caption{Model of a thermoelectric generator with ideal thermal contacts to heat reservoirs (top) and nonideal thermal contacts (bottom).}
	\label{fig:figure1}
\end{figure}

For simplicity, but without loss of generality, we consider a thermoelectric generator with constant parameters, i.e., independent of the local temperature. The TEG, depicted in Fig.~\ref{fig:figure1}, is characterized by a Seebeck coefficient $\alpha$, a thermal conductivity $\kappa$ and an electrical conductivity $\sigma$. From a global view point on the system, it is more convenient to consider its electrical resistance $R$ defined as $R = l / (\sigma A)$, with $A$ being the section of the TEG, and $l$ its length, and its thermal conductance $K_0$ defined as $K_0 = A \kappa / l$. To produce power, the TEG is placed between two heat reservoirs at temperatures $T_{\rm h}$ and $T_{\rm c}$ respectively, with $T_{\rm h} > T_{\rm c}$. The temperature difference across the TEG is denoted $\Delta T$ and the average temperature is given by $\overline{T} = (T_{\rm h}+T_{\rm c})/2$; for convenience we assume that $\overline{T}$ retains the same value during operation.

The potential performances of a TEG may be evaluated using its figure of merit $Z\overline{T}$, where $Z= \alpha^2 \sigma/\kappa$ or, identically, $Z= \alpha^2 /(RK_0)$ \cite{Ioffe1957}. Indeed $Z\overline{T}$ is tightly linked to the TEG maximum efficiency $\eta_{\rm max}$ since in the classical system where $\Delta T$ is assumed to be constant one may find:
\begin{equation}
\eta_{\rm max} = \frac{\Delta T}{T_{\rm h}} \frac{\sqrt{1+Z\overline{T}}-1}{\sqrt{1+Z\overline{T}}+T_{\rm c}/T_{\rm h}}.
\end{equation}
\noindent This maximal efficiency is reached through an appropriate choice of the electrical load resistance $R_{\rm load}$: In this case $m = \sqrt{1+Z\overline{T}}$, $m$ being the ratio between $R_{\rm load}$ and $R$ \cite{Ioffe1957}.

One should note however that the previous result only holds for $\Delta T$ set as constant. In order to examine TEG optimization in other cases, we must express the dependence of both output power and efficiency on the thermal and electrical constraints. We thus use the equivalent thermal resistance of the TEG, defined as 
\begin{equation}\label{ThetaTE}
\Theta_{\rm TE}= \Theta_{0} \frac{1}{1 +\alpha \overline{T} I/K_0\Delta T},
\end{equation}
\noindent with $\Theta_{0} = 1/K_0$, to explicitly express the relation between the electrical and the thermal parts of the system. Indeed, in the previous expression, one may notice that $\Theta_{\rm TE}$ depends on the electrical current $I$ flowing in the system. The thermal resistance $\Theta_{0}$ thus appears as the thermal resistance of the system for open electrical circuit condition. The influence of the electrical current on the thermal properties of the TEG is due to the contribution of a \emph{convective thermal flux}, i.e., associated to the global displacement of the electrons \cite{Apertet2012, Apertet2012JPCS}. This contribution may be seen as a genuine footprint of thermoelectric transport. Since, in first approximation, the incoming heat flux $q_h$ is given by
\begin{equation}\label{qh}
q_h = \frac{\Delta T}{\Theta_{\rm TE}},
\end{equation}
\noindent it is possible to provide a simple relationship between the different parameters of the system. We also need to use the relation between efficiency $\eta$ and output power $P$: $P = \eta q_h$.

This model may be extended to consider finite thermal resistances between the TEG and the heat reservoirs (see bottom panel of Fig.~\ref{fig:figure1}). The thermal resistance on the hot(cold) side of the TEG is labeled $\Theta_{\rm Hx,h}$ ($\Theta_{\rm Hx,c}$). The total thermal resistance of the contacts is defined as $\Theta_{\rm Hx} = \Theta_{\rm Hx,h} + \Theta_{\rm Hx,c}$.

\subsection{On the choice of the thermal constraints}

In the traditional approach of the TEG optimization, one considers that the temperatures of the heat reservoirs, to which the TEG is connected, are constant. In this case, the incoming heat flux $q_h$ varies when the electrical load resistance is changed. This situation may be seen as the equivalent of a \emph{thermal Th\'evenin model}: the potential difference, here the temperature difference, is fixed while the current changes depending on the load. As already stressed, this model may be extended to deal with more realistic situation where non ideal thermal contacts are considered (see Fig.~\ref{fig:figure1}). The thermal resistance of the contacts may thus be seen as the internal resistance of the Th\'evenin generator.

Although this is a less common approach, it also possible to assume that the incoming thermal flux $q_h$ is constant (see e.g. \cite{Snyder2006, Baranowski2014, Raag1969, Okhotin1972}). This viewpoint should thus be related to a \emph{Norton model} of a thermal circuit. In this case, the temperature difference $\Delta T$ varies as the electrical load resistance changes. While this behavior is also observed when nonideal thermal contacts are taken into account in the \emph{Th\'evenin model} (see e.g. \cite{Freunek2009, Yazawa2011, Apertet2012}), its origin is quite different. Indeed with a \emph{Norton model}, it is not possible to account for the impact of the thermal contacts on the performances of the TEG since a resistance in series with an ideal current generator imposing a constant $q_h$ has no effect whatsoever on the rest of the circuit. Hence the \emph{Norton model} seems inappropriate to deal with practical applications.

\subsection{Optimization with constant incoming heat flux}

Since the \emph{Norton model} of the TEG thermal environment differs from the classical \emph{Th\'evenin model}, the optimal working conditions in this case must be derived rigorously. For both models, the output power reads: 
\begin{equation}
P = \frac{\alpha^2 \Delta T^2}{R}\frac{m}{(m+1)^2}.
\end{equation}
\noindent When $\Delta T$ is constant, i.e. for the \emph{perfect Th\'evenin model}, one immediately notices that the maximization condition is $m = 1$. However, for the \emph{Norton model}, it is mandatory to explicit the dependence of $\Delta T$ on the ratio $m$ associated to the electrical load condition. To do so, we combine Eqs.~(\ref{ThetaTE}) and (\ref{qh}) and substitute the electrical current $I$ by the following expression obtained from the electrical part of the TEG:
\begin{equation}
I = \frac{\alpha \Delta T}{R \left(m+1\right)}
\end{equation}
\noindent The temperature difference across the TEG thus becomes:
\begin{equation}
\Delta T = \frac{q_h}{K_0}\frac{m+1}{m+1+Z\overline{T}}
\end{equation}
\noindent where the ratio $m$ is the only variable as all other parameters are fixed.

The equivalent thermal resistance of the TEG in this case reads:
\begin{equation}
\Theta_{\rm TE}  = \frac{m+1}{K_0 \left(m+1+Z\overline{T}\right)}
\end{equation}
\noindent and the output power is then given by:
\begin{equation}
P = \frac{Z q_h^2}{K_0}\frac{m}{(m+1+Z\overline{T})^2}.
\end{equation}
\noindent Finally, for the \emph{Norton model}, the power maximization condition simply is:
\begin{equation}
m_{\rm opt} = 1+Z\overline{T}.
\end{equation}
\noindent Since the TEG efficiency $\eta$ is related to the output power $P$ through $P = \eta q_h$, this condition also maximizes efficiency.

\begin{figure}
	\centering
		\includegraphics*[width=0.47\textwidth]{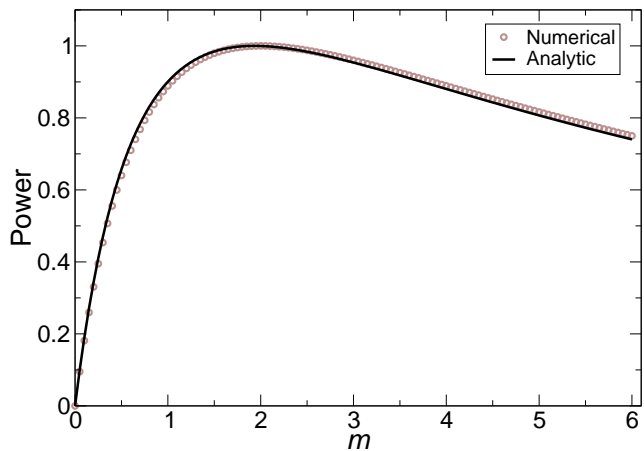}
	\caption{Normalized output power as a function of the ratio $m$.}
	\label{fig:figure2}
\end{figure}

To check the validity of the previous analytical expression, we use a numerical model of the TEG where the incoming thermal flux $q_h$ is no longer approximated using Eq.~(\ref{qh}). Instead we use the exact expression given in \cite{Ioffe1957}:  
\begin{equation}
q_h = \alpha T_{\rm hot} I + K_0\Delta T - \frac{1}{2}R I^2
\end{equation}
\noindent As discussed in Ref.~\cite{Apertet2012}, the approximated expression for $q_h$ remains valid as long as the temperature difference $\Delta T$ is negligible compared to the average temperature $\overline{T}$ of the TEG. 
The parameters for TEG are set to the following values: $\overline{T} = 300~K$, $Z\overline{T} = 1$, $K_0 = 2.5$~mW/K, $R = 4.8~$m$\Omega$, and $\alpha = 200 ~\mu$V/K. These parameters are consistent with the properties at room temperature of a particular bismuth telluride compound, (Bi$_{0.25}$Sb$_{0.75}$)$_2$Te$_3$, one of the most efficient thermoelectric materials \cite{Yamashita2003}. The incoming thermal flux, arbitrarily chosen, is set at $200$~mW. Figure~\ref{fig:figure2} shows the evolution of the normalized output power $P/P_{\rm max}$ as a function of the ratio $m$. The comparison between the analytical and numerical models demonstrates that the approximated expression for $q_h$ allows to correctly describe the TEG behavior, even if the temperature difference $\Delta T$ reaches $60~K$ at the maximum power working condition. The curves thus confirm that for Norton model the optimal condition is $m_{\rm opt} = 1+Z\overline{T}$. Note that this condition was already given by Okhotin et al. in the Russian version of Ref.~\cite{Okhotin1972} though a different approach was used.

This result forces to reconsider the claims made by Baranowski et al. in Ref~\cite{Baranowski2014}: In that article, the authors indeed compare two thermoelectric generators with identical incoming heat fluxes but with different $m$ values. They demonstrate that a TEG with $m=\sqrt{1+Z\overline{T}}$ gives more electrical power than a TEG with $m=1$. They then conclude that the produced power is maximized for $m=\sqrt{1+Z\overline{T}}$. Thanks to our analysis, it is now possible to state that the power is indeed increased when shifting from the condition $m=1$ to the condition $m=\sqrt{1+Z\overline{T}}$ but that the power is actually maximized for $m = 1+Z\overline{T}$ due to the hypothesis of constant incoming heat flux.
%We also stress that the purpose of this comparison, associated to the Norton modelization since $q_h$ is fixed, is to demonstrate that the model presented in Ref.~\cite{Baranowski2013} is accurate whereas this model is associated to constant temperature difference, i.e., to the Th\'evenin model. It is interesting to note that Ragg already anticipated this possible confusion between the two models and warned against it in Ref.~\cite{Raag1969}.

\subsection{On the simultaneity of efficiency and power maximization}

\begin{figure}
	\centering
		\includegraphics*[width=0.47\textwidth]{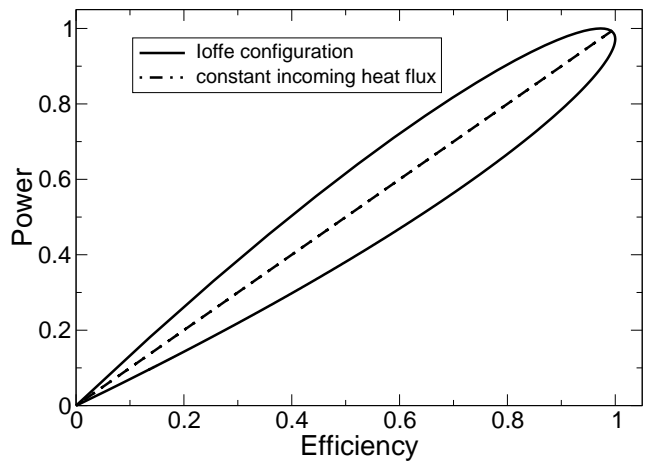}
	\caption{Normalized output power ($P/P_{\rm max}$) as a function of the normalized efficiency ($\eta/\eta_{\rm max}$) for the two kinds of TEG. Each curve is described by varying the load resistance $R_{\rm load}$, i.e. the ratio $m$.}
	\label{fig:figure3}
\end{figure}

Recently, there has been some discussions on the possibility to reach simultaneously working conditions leading to both output power and efficiency maximizations \cite{Apertet2014, Baranowski2014}. The distinction between \emph{Th\'evenin} and \emph{Norton models} for the TEG thermal environment allows to shed light on this issue: Indeed, it seems clear that, while for the \emph{perfect Th\'evenin model} as considered by Ioffe \cite{Ioffe1957}, optimal working conditions differ for efficiency: $m_{\rm opt} = \sqrt{1+Z\overline{T}}$, and output power: $m_{\rm opt} = 1$; these two optimization targets are maximized simultaneously for the \emph{Norton model} since in this case they are proportional to each other. This difference is highlighted on Fig.~\ref{fig:figure3}, where the lobe shape of the $P$ vs. $\eta$ curve for the \emph{perfect Th\'evenin model}, shows that the points for maximum efficiency and maximum output power are distinct. Note that increasing $Z\overline{T}$ leads to open this loop and hence to further seclude both points.

The confusion between efficiency and power maximization may also have been triggered by the results concerning models of TEGs with non ideal thermal contacts: The condition for power maximization in such a case, and when thermal impedance matching is satisfied, is $m=\sqrt{1+Z\overline{T}}$ \cite{Freunek2009, Apertet2012, Yazawa2012, McCarty2012, Gomez2013}, which appears to be traditionally associated to efficiency maximization for a TEG with \emph{ideal} thermal contacts. However, it is mandatory to keep in mind that the systems are different. Deriving the conditions of efficiency maximization (see e.g. Ref.~\cite{Apertet2012}), it is then possible to demonstrate that the two conditions are distinct just as in the case of the \emph{perfect Th\'evenin model}, even if the difference between them depends on $Z\overline{T}$ and vanishes when the figure of merit vanishes. Interestingly, in Ref.~\cite{Gomez2013}, Gomez and coworkers report numerical simulations of TEGs with nonideal thermal contacts. They find a 5\% difference between the $m$ value for power maximization and the $m$ value for efficiency maximization. While this discrepancy may be interpreted as a consequence of approximations within the model, it seems more probable that this small difference is related to the low $Z\overline{T}$ value of their system since their results are in good agreement with the analytical expressions obtained in Ref.~\cite{Apertet2012}.

\section{\label{procedure} Simplified design rules to maximize output power of a TEG}

We now turn to the practical case of the design of a thermoelectric generator described with the constant parameter model. As stressed in Ref.~\cite{Baranowski2014}, a common way to tackle this issue is to \emph{frame the design problem in terms of a given waste heat source, for which a heat exchanger would be the first system component selected, thus fixing $\Theta_{\rm Hx}$}, the total thermal conductance of the contacts. So, with this constrained thermal environment, the power optimization of the generator may be performed in 3 steps:\\

 \noindent 1. \emph{Choice of the best material}\\
One has to choose the material with the maximum figure of merit $Z\overline{T}$ available in the desired temperature range.

 \noindent 2. \emph{Determination the module length using the thermal impedance matching}\\
This step concerns the geometry of the module. It has been shown that a thermal impedance matching is mandatory to maximize output power, i.e., that the thermal resistance of the module $\Theta_{\rm TE}$ should be equal to the global thermal resistance $\Theta_{\rm Hx}$ of the contacts: $\Theta_{\rm TE} = \Theta_{\rm Hx}$ \cite{Stevens2001}. As a first approximation, one could identify the thermal resistance of the module $\Theta_{\rm TE}$ to its thermal resistance under open electrical circuit condition $\Theta_{\rm 0}$. However, as stressed in Eq.~(\ref{ThetaTE}), the electrical current $I$ flowing through the TEG modifies the global thermal resistance $\Theta_{\rm TE}$. Consequently, in order to fulfill the thermal matching condition, one must anticipate the value of the electrical current during operation at maximum power. With such a consideration, the thermal matching condition becomes:
\begin{equation}\label{eq:thermalmatching}
\Theta_{\rm 0} = \Theta_{\rm Hx} \sqrt{1 + Z\overline{T}}
\end{equation}
as demonstrated in Refs.~\cite{Freunek2009, Yazawa2011, Apertet2012, Yazawa2012, McCarty2012}. It is then straightforward to find the corresponding optimal value for the TEG length \cite{Yazawa2011, Yazawa2012}:
\begin{equation}
l_{\rm opt} = \kappa A \Theta_{\rm Hx} \sqrt{1 + Z\overline{T}}.
\end{equation}
\noindent Note that the larger the surface $A$ of the TEG, the larger the output power $P$. Since this surface is not constrained by the thermal environment, $A$ should be maximized in the limit of the available space.
\\

3. \emph{Determination of the electrical load using the electrical impedance matching}

Finally, the electrical working condition must be set. In order to maximize power, it is mandatory to satisfy electrical impedance matching, i.e., $R_{\rm load} = R_{\rm TE}$ where $R_{\rm TE}$ is the electrical resistance of the TEG. In the case of ideal thermal contacts, $R_{\rm TE}$ simply is the resistance $R$. However, when non ideal thermal contacts are considered, just as an additional term appears for the thermal resistance of the TEG, an additional term appears in $R_{\rm TE}$, reflecting the interdependence of the thermal and electrical parts of the TEG \cite{Apertet2012}. We have demonstrated that taking account of this additional term, the electrical load resistance maximizing output power is then
\begin{equation}
R^{\rm opt}_{\rm load} = R \left(1 + \frac{Z\overline{T}}{1+\Theta_{\rm 0} / \Theta_{\rm Hx}}\right).
\end{equation}

\noindent It is interesting to notice that this electrical condition depends on the thermal environment of the TEG through the appearance of $\Theta_{\rm Hx}$. Considering that $\Theta_{\rm 0}$ have been set correctly during the second step in order to satisfy the thermal impedance matching condition, Eq.~(\ref{eq:thermalmatching}), the previous equation becomes
\begin{equation}
R^{\rm opt}_{\rm load} = R \sqrt{1 + Z\overline{T}}.
\end{equation}
\noindent It then appears clearly that this condition is valid only when thermal impedance matching condition is also met as we have already highlighted. If, however, the range of available variation range for $R_{\rm load}$ does not allow to satisfy this condition, it is possible to use a DC-DC converter in order to modify the apparent electrical load resistance seen from the TEG \cite{Molan2011}.
\\

Once both thermal and electrical impedance matching conditions are satisfied, the output power is then \cite{Yazawa2011, Apertet2012, Yazawa2012}:
\begin{equation}
P_{\rm max} = \frac{Z\overline{T}}{\Theta_{\rm Hx} \left(1+\sqrt{1+Z\overline{T}}\right)^2} \frac{(\Delta T')^2}{4\overline{T}}.
\end{equation}
\noindent where $\Delta T'$ is, in this case, the temperature difference between the hot and cold heat reservoirs. This equation justifies the choice of a material with the best figure of merit $Z\overline{T}$ rather than the best power factor $\alpha^2 \sigma$ even if one seeks power maximization. 

\section{Conclusion}

We have highlighted the importance of the assumptions regarding the thermal environment of a TEG on its optimization. We distinguished the \emph{Th\'evenin model}, for which the temperature difference between the heat reservoirs is supposed to be constant, from the \emph{Norton model}, for which the incoming heat flux is supposed to be constant. Using this distinction, we discuss the possibility to reach simultaneously both efficiency and power output maximization. We have demonstrated that this situation is only feasible for the \emph{Norton model}. In this case, it was shown that the optimal working condition is obtained for $m = 1 +Z\overline{T}$. Finally, we provided a simple practical procedure to design a TEG aiming at maximum output power. We hope that this article will help to clarify ongoing debates \cite{Baranowski2014, Apertet2014} on TEG optimization.

\begin{acknowledgments}
We are pleased to thank Dr. K. Zabrocki and one of the Referees for bringing to our attention respectively Ref.~\cite{Raag1969} and Ref.~\cite{Okhotin1972}.
\end{acknowledgments}

\end{document}